\documentstyle[aps]{revtex}
\newcommand{\dpar}[2]{\frac{\partial #1}{\partial #2}}
\def\be{\begin{equation}}
\def\ee{\end{equation}}
\def\ba{\begin{eqnarray}}
\def\ea{\end{eqnarray}}

\def\tel{\tau_{\rm el}}
\def\Pext{\Phi_{\rm ext}}
\def\bl{\bar{\lambda}}
\def\lkin{\ell}
\def\ind{{\cal L}}
\def\ltot{{\cal L}_{\rm tot}}
\def\vcr{v_{\rm cr}}
\def\Scr{S_{\rm cr}}
\def\a{\alpha}
\def\b{\beta}
\def\k{\bf{k}}
\def\f{f_{\bf{k}}}
\def\tE{{\tilde E}}
\input epsf
\begin{document}
\draft
\twocolumn[\hsize\textwidth\columnwidth\hsize\csname
@twocolumnfalse\endcsname
\title{Numerical study of induced vortex tunneling}
\author{C. D. Bass and S. Khlebnikov}
\address{
Department of Physics, Purdue University, West Lafayette, 
IN 47907, USA}
\maketitle
\begin{abstract}
Tunneling of vortex-antivortex pairs across
a superconducting film can be controlled via inductive coupling of
the film to an external circuit. We study this process numerically in 
a toroidal 
film (periodic boundary conditions in both directions) by using the dual
description of vortices, in which they are represented by a fundamental 
quantum field. We compare the results to those obtained in the
instanton approach.
\end{abstract}
\vskip2pc]

\section{Introduction}
Persistent-current superconducting devices,
in which the basis states are characterized by different values of 
the enclosed flux, are interesting physical systems in their own right
and are also promising candidates for applications to quantum memory.
The possibility to form quantum superpositions
of macroscopic flux states has been demonstrated in experiments 
with SQUIDs \cite{1,2}.

Once reliable storage of quantum superpositions is achieved, it becomes
necessary to consider possible mechanisms for reading and writing operations
and for assembling several such individual units (qubits) into a scalable
quantum computer. In SQUIDs, various proposals have exploited the existence
of a potential barrier between two stable basis states and have involved
manipulation of the barrier itself, use of tunneling effects, or a 
controlled excitation over the barrier \cite{tipping,deex}.

\begin{figure}
\leavevmode\epsfxsize=2.5in\epsfbox{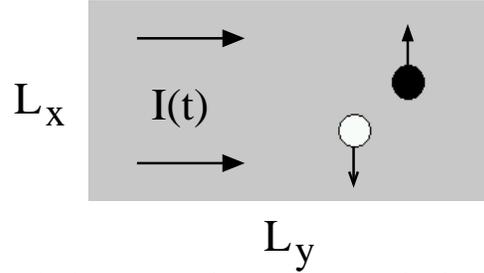}
\caption{A schematic of vortex tunneling induced by a pulse of
supercurrent $I(t)$. The circles denote a vortex and an antivortex.}
\label{fig:film}
\end{figure}

In this paper, we examine a model which describes tunneling of vortices
(short Abrikosov flux lines) in a ring of thin superconducting film. We consider 
a scheme in which a pulse of supercurrent suppresses superconductivity, thus lowering 
the potential barrier and inducing tunneling, see Fig. \ref{fig:film}. 
Because this process
changes the flux enclosed by the ring, it can be used to form arbitrary
superpositions of the basis states. We will see that in a suitable geometry, it
is possible also to independently control the energy bias between the 
basis states, similarly to how it is done in SQUIDs.
On the other hand, this device does not contain 
any Josephson junctions, thereby avoiding dissipation due to various 
fabrication issues, such as defects in the insulating barrier. 

Motivated by these observations, we consider a simplified model, convenient for
numerical work, in which the film has periodic boundary conditions in both directions, 
forming a torus. The suppression of superconductivity and the biasing flux are
represented in this model by two independent parameters: the vortex pair-production
frequency $M(t)$, and the driving force $F(t)$.

For applications to quantum computing, of main interest is the adiabatic
regime, when there is very little residual excitation left in the system
after the pulse (in other words, no real, as opposed to virtual, vortex
pairs are produced). If this condition were not satisfied, the remaining
vortices would be easily ``detected'' by the environment 
(e.g. by electrons at the vortex cores), and that would result in rapid 
decoherence. Thus, we envision a situation when a virtual 
vortex and an antivortex are created, say, on the inside of the torus, 
transported along the opposite semicircles to the outside, and 
annihilated there, almost without a trace.

In thin films, the elastic mean-free time of electrons is short \cite{Fuchs}; 
for a film of thickness $d$ we use
$\tel = 2d / v_F$. This results in a strong suppression of the Magnus force
on the vortex and a relatively small friction \cite{KK,Volovik,micro,FGLV}. 
In addition, and in contrast
to motion of real vortices, the density of normal electrons at the vortex
core during tunneling is a variational parameter, which adjusts itself to
maximize the tunneling rate. This leads to further cancellation of the Magnus
force and a reduction in the inertial mass of the vortex.

There are two theoretical approaches to induced vortex transport. 
One is based on instantons, solutions to the Euclidean equations of motion.
Using the expressions \cite{KK,micro} for the friction caused by the core
fermions, we find that it results in the following contribution to
the Euclidean action:
\be
S_f = \pi \omega_0 \tel n_e L_x^2 d \; ,
\label{Sf}
\ee
where $n_e = k_F^3 / 3\pi^2$ is the electron density, $L_x$ is the width of the
film, and $\hbar \omega_0 \sim \Delta^2 / \epsilon_F$. Using $k_F = 1$ \AA$^{-1}$,
$2\omega_0 = 10^{-8} k_F v_F$, $L_x = 1\mu$m, and $d=4$ nm, we obtain
$S_f = 170$. By itself, this is not a small number, but the crucial
point is that $S_f$ depends quadratically on the gap $\Delta$. So, when we suppress
$\Delta$ by a pulse of current, we also reduce $S_f$. In fact, the dependence
of $S_f$ on $\Delta$ is precisely the same as that of the vortex 
pair-production frequency $M$. So, in what follows we simply include the effect 
of the friction in our definition of $M$.

The driving force $F$, due to the energy bias between the basis states, 
can be viewed as a Lorentz force 
caused by an effective electric field, $E'$, in the $y$ direction. Then, the 
average vortex  current due to tunneling is obtained in the instanton approach 
as \cite{tran}
\ba
 \langle J_x \rangle & \sim & e^{-S_0+i\tE L_x} - e^{-S_0-i\tE L_x}
\nonumber \\
& & \sim  e^{-S_0} \sin \tE L_x \; ,
\label{cur0}
\ea
$\tE = (d/4e) E'$, and $e$ the magnitude of the electron
charge. In (\ref{cur0}),
the first term is due to instantons, and the second to anti-instantons;
$S_0 = M L_x/ c_1$ is the real part of the instanton action, and $c_1$ is the
limiting speed of vortex motion.

The second method is entirely real-time. Vortices are described by a fundamental
quantum field, and a nonzero average vortex current comes out as a result
of the discreteness of field modes. 
The periodicity in $\tE$ has been confirmed analytically in this approach \cite{tran},
provided the vortex-antivortex potential can be replaced by its average and included
as an additional contribution to $M$. 
The second principal effect seen in (\ref{cur0})---the $\exp(-M L_x/ c_1)$ 
dependence---has been confirmed only for the case of small $\tE$,
$\tE \ll 2\pi / L_x$. It is of interest to develop this approach further,
so that it can be applied also to cases with large $\tE$ and a non-trivial
potential. The present paper addresses, via numerical integrations, the first part 
of this program. 

In practice, it may be easier 
to fabricate a thin strip than a thin cylinder. The field-theoretical method will
apply to that case as well, provided one can establish the boundary
conditions for the vortex field at the edges of the strip.

The paper is organized as follows.
In Sect. \ref{sect:cont} we discuss how one can independently control the vortex mass 
and the driving force in a thin superconducting ring.
In Sect. \ref{sect:real}, we discuss the real-time description of vortex tunneling.
Numerical results are presented in Sect. \ref{sect:num}, and a summary
in the concluding Sect. \ref{sect:concl}.

\section{Control of current and flux in thin superconductors} \label{sect:cont}
Consider a uniform superconducting ring inductively coupled to an external 
circuit. Suppose the order parameter winds $n$ times around the ring.
Then, the London current can be expressed through the flux $\Phi$
enclosed by the ring as
\be
I = - c (\Phi - n\Phi_0) / \lkin \; .
\label{I}
\ee
Here $\Phi_0$ is the flux quantum, and $\lkin$ is
the ``kinetic'' inductance:
\be
\lkin = \frac{mc^2 L_y}{e^2 n_s S} = 4\pi \bl^2 \frac{L_y}{S} \; ,
\label{Lkin}
\ee
$n_s$ is the density of superconducting electrons, $S$ is the cross-sectional
area, and $L_y$ is the length of the ring. 

We have introduced the London
penetration depth $\bl$, and because superconductivity can be
deliberately suppressed by some means, $\bl$ is in general different from the
unperturbed penetration depth $\lambda$. Recall also that in a thin film $\bl$
determines the strength of the London current, but not the screening length of
the magnetic field \cite{Pearl}.

The flux $\Phi$ in (\ref{I}) is the total flux, which consists of the flux created 
by the external circuit and that created by the current $I$ itself: 
\be
\Phi = \Pext + {1\over c} \ind_0 I \; ,
\ee
where $\ind_0$ is the ordinary inductance of the ring. Using this together with
eq. (\ref{I}), we can express the supercurrent through $\Pext$:
\be
I = -c \frac{\Pext - n\Phi_0}{\lkin + \ind_0} \; .
\label{cur}
\ee
Even though $\lkin$ and $\ind_0$ enter eq. (\ref{cur}) symmetrically, there is
an essential difference between them, since $\ind_0$ depends only on the geometry
and in this  sense is a constant, while $\lkin$ depends on $n_s$.

So, there are two distinct limits of eq. (\ref{cur}). 
If $\lkin \ll \ind_0$, i.e.,
the cross-sectional area $S$ is large enough, then according to 
(\ref{cur}) $\Pext$ controls the current. 
On the other hand, if $\lkin \gg \ind_0 $, $\Pext$ determines
only the product $\lkin I$, i.e., the ratio $I / n_s$.

In the second, thin-ring, regime, by a slow variation of $I/n_s$ 
we can smoothly
change the order parameter $\psi$ from a large initial value, for which vortex
tunneling will be strongly suppressed, to some much smaller values that 
allow tunneling, and then back to the initial state. For small, slowly changing
$\psi$, this can be seen directly from the Ginzburg-Landau (GL) equation:
\be
\left( \frac{j}{e n_s \vcr} \right)^2 - 1 + {b \over |a|} 
|\psi|^2 = 0 \; ,
\label{GL}
\ee
where $j$ is the current density, $\vcr = \hbar/2m\xi$ is a critical velocity
($\xi$ is the coherence length), and $a$, $b$ are GL parameters. 
If quantum 
coherence can be preserved during this process, such a device would 
be reliable quantum memory. 

Now, $\ind_0\approx 2 L_y \ln(L_y/L_x)$, so the crossover between the thick and 
thin-ring cases occurs at
\be
\Scr \sim 2\pi \bl^2 \ln^{-1}(L_y/L_x) \; .
\ee
Assuming that the logarithm is of order unity, and using an unperturbed value 
$\bl = \lambda = 150$ nm, we obtain
\be
\Scr \sim 1.4\times 10^5 {\rm nm}^2 \; .
\label{Scr}
\ee
When $n_s$ (which is proportional to $|\psi|^2$) is suppressed to allow
tunneling, $\bl$ and, consequently, $\Scr$ grow, so if the condition
$S < \Scr$ held in the initial state, it would hold even better during tunneling.

According to this estimate, if the ring is made from a thin film, it does not 
have to be particularly narrow to achieve the thin-ring condition $S< \Scr$. 
For example, for thickness $d = 4$ nm, the
estimate (\ref{Scr}) allows for widths as large as $10$ $\mu$m. 

As a specific example of how $n_s$ can be suppressed by a pulse of current, consider
the double-arm geometry shown in Fig. 1. 
It is convenient to imagine that the wire carrying the constant current $I$ is 
closed at a large distance,
so that the device can be viewed as a superposition of two closed circuits, with
currents $I_1$ and $I= I_1+I_2$. In addition to kinetic inductances $\lkin_1$,
$\lkin_2$, the circuits have self-inductances $\ind_{11}$, $\ind_{22}$
and a mutual inductance $-\ind_{12}$. We can also define 
$\ind_1 \equiv \ind_{11} - \ind_{12}$ and $\ind_2 \equiv \ind_{12}$.
Consider regime when the inductance of arm 1 is mostly kinetic, $\lkin_1 \gg \ind_1$,
while that of arm 2 is mostly ordinary, $\lkin_2 \ll \ind_2$. Then, the currents
in the arms are
\ba
I_1 & = & \frac{1}{\ltot} [\ind_2 I - c(\Pext - n \Phi_0)] \; ,
\label{I1}
\\
I_2 & = & \frac{1}{\ltot} [\lkin_1 I + c(\Pext - n \Phi_0)] \; ,
\label{I2}
\ea
where $\ltot = \lkin_1 + \ind_2$. Suppose further that $\lkin_1 \gg \ind_2$.
In this case, we see from (\ref{I1}) that the current $I$ controls the
parameter $I_1 \lkin_1$, which according to (\ref{GL}) determines how close the first
arm is to criticality.

\begin{figure}
\leavevmode\epsfxsize=1.7in\epsfbox{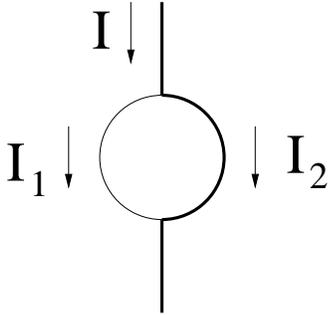}
\caption{A double-arm device, in which suppression of superconductivity in the
weaker (first) arm is controlled by an external current $I$, while a biasing flux 
controls the energy difference between two flux states.}
\label{fig:dev}
\end{figure}

The inductive energy of the double-arm device is
\be
{\cal E} = {1\over 2\ltot} (\Pext - n \Phi_0)^2 + \epsilon(I^2) \; ,
\label{ene}
\ee
where $\epsilon(I^2)$ is independent of $n$ and the external flux. If the device
is biased by half a flux quantum, $\Pext = \Phi_0 / 2$, the energy (\ref{ene})
has two degenerate minima at $n= 0,1$. If $\Pext$ deviates from half-quantum
by a small amount $\Delta \Pext$, the energy difference between the two minima
is $\Delta {\cal E} = \Phi_0 \Delta \Pext / \ltot$.
This results in an additional force 
\be
F=\Delta {\cal E} / L_x 
\label{F}
\ee
acting on a vortex as it tunnels across arm 1. 
Note that unlike the case of a single
current-biased superconducting wire \cite{GF}, this force is not related to the total
current $I$ but is an independently controlled parameter. The only restriction
is that $|\Delta {\cal E} |$ should not exceed the energy $2M$ of vortex pair production,
so that no real vortices are able to nucleate.

\section{Real-time description of vortex tunneling} \label{sect:real}
Motivated by the arguments of the preceding sections we consider a model
of vortex tunneling, in which the main force acting on the vortex is the
driving force (\ref{F}). The requisite suppression of the order parameter is
achieved by some independent means, such as a pulse of external current in the
double-arm device. 

In the real-time approach, vortices are described by a quantum field $\chi$
\cite{Ao}, which in our case obeys the equation of motion
\ba
\left[\partial_t + {i\over \hbar} U(x,t) \right]^2 \chi 
- c_1^2 \left[ \partial_x - i\frac{d}{4e} E(t)\right]^2\chi & & \\
 - c_1^2 \partial_y^2\chi + M^2(t)\chi & = & 0 
\; .
\label{eqm}
\ea
Here $d$ is the thickness of the film, $e$ is the magnitude of the electron
charge, and $E(t)$ is a time dependent electric field produced by the changes
in $I$ and $\Pext$. The speed $c_1$ is the limiting speed of vortex motion:
in the second-quantized description (\ref{eqm}) it plays the role analogous
to the speed of light in relativistic quantum theory. 

The driving force is represented by the potential
\be
U(x, t) = -F(t) x \; .
\ee
Such an explicit $x$-dependence in the equation is inconvenient for numerical
work. However, it is possible to make a transformation of the field $\chi$, 
similar to a gauge transformation, so that the force disappears from the
first term in (\ref{eqm}) and appears instead as an addition to the electric
field $E$:
\ba
\chi & \to & \chi\exp [{i\over \hbar} \int_{-\infty}^t U(x, t') dt' ] \; , 
\label{tran_chi} \\
U & \to & 0 \; , \label{tran_U} \\
E & \to & E' = E - \frac{4e}{\hbar d} \int_{-\infty}^t F(t') dt' \label{tran_E} \; .
\ea
In what follows, we will use the same notation $\chi$ for the transformed field
as we used for the original one.

In general, the transformations (\ref{tran_chi})--(\ref{tran_E}) lead to one spurious 
effect. Imagine that $F(t)$ starts from zero at $t=-\infty$, 
goes through
nonzero values near $t = 0$ and then back to zero at $t=\infty$. Then, 
according to eq. (\ref{eqm}), $U$ has no effect at $t \to \infty$, while according to 
(\ref{tran_E}) the correction to $E$ is still nonzero (and proportional to the
integral of $F$).
Fortunately, owing to the periodic dependence of vortex transport on $\tE$,
cf. eq. (\ref{cur0}), this correction is immaterial provided
$F$ satisfies a quantization condition:
\be
 \int_{-\infty}^\infty F(t) dt = \frac{2\pi \hbar n'}{L_x} \; ,
\label{quant}
\ee
where $n'$ is an integer. Only in this case the problem obtained by the
transformations (\ref{tran_chi})--(\ref{tran_E}) is equivalent to the original
problem (\ref{eqm}).

The transformed equation has no explicit dependence on $x$ and can be solved
by the mode expansion
\be
\chi({\bf{x}},t) = \sqrt{\hbar}
\sum_{\k} \left[ \a_{\k} {\f} (t) + \b^{\dagger}_{-\k} {\f}
^*(t) \right] e^{i {\k} \bf{x}} \; ,
\label{exp}
\ee
where ${\k} = (k_x, k_y)$, $\a$ and $\b$ are annihilation operators for vortices
and antivortices, respectively, and ${\f} (t)$ are the mode functions that 
take into account the time dependence of $E'(t)$ and of the vortex ``mass'' 
$M(t)$. We have assumed that the vortex field has periodic boundary conditions
in both directions.

Substituting the expansion (\ref{exp}) into the field equation, 
we obtain the equation for the mode functions:
\be
\ddot{\f} (t) + \omega_{\k}^2(t) \f(t) = 0
\label{eqf}
\ee
where
\be
\omega_{\k}^2(t) = c_1^2 k_y^2 + c_1^2 [k_x - \tE(t)]^2 
+ M^2(t) \; ,
\label{ome}
\ee 
and $\tE =(d/4e) E'$. We consider the case of effectively zero
temperature, when there are no vortices in the initial state. So, eq. (\ref{eqf}) is
solved with the vacuum initial conditions
\be
f_{\k}(t \to -\infty) = [2\omega^{(0)}_k V]^{-1/2}
\exp[-i \omega^{(0)}_k (t - t_i)] \; ,
\label{init}
\ee
where $V$ is the two dimensional volume of the film, $t_i$ is some initial moment of time,
$\omega^{(0)}_k= [c_1^2 k^2 + M_0^2]^{1/2}$, and $M_0=M(t\to -\infty)$.

Once a  solution to the initial problem (\ref{eqf})--(\ref{init}) is
available (e.g., from a numerical integration), one can obtain various
quantities of interest as averages over the vacuum of the operators $\a$ and $\b$.
In what follows, we consider three such quantities: the average vortex current,
the energy, and the vortex occupation numbers, all as functions of time.

Only the $x$ component of the average vortex current is nontrivial. It can
be found by averaging the operator expression
\be
\frac{\hbar}{c_1^2} J_x(t) = 
-i \chi^{\dagger} \partial_x \chi + i (\partial_x \chi^{\dagger}) \chi 
- 2\tE \chi^{\dagger} \chi
\ee
over the vacuum of $\a$ and $\b$, to obtain
\be
\langle J_x(t) \rangle = \sum_{k_y} q(t, k_y) \; ,
\label{J}
\ee
where
\be 
q(t, k_y) = 2 c_1^2 \sum_{k_x} [k_x - \tE(t)] |{\f}|^2 \; .
\label{q}
\ee
If we integrate (\ref{J}) over time, we will obtain the average vortex number
transported in the $x$ direction per unit length in 
the $y$ direction during the entire pulse. This is a convenient
measure of vortex transport.

At zero temperature, and for a sufficiently adiabatic pulse, the only source of
vortex transport is vortex tunneling. In the real-time formalism, the 
corresponding contribution to the current (\ref{J}) appears as a result of
the discreteness of modes. It is exponentially suppressed with
$L_x$ and should reproduce the result (\ref{cur0}) of the instanton approach.

We should be careful, however, about the ultraviolet regularization of
eq. (\ref{q}). A sharp momentum cutoff is not adequate for our purposes, 
especially since we need a regulator that would preserve the symmetry under the
transformation (\ref{tran_chi})--(\ref{tran_E}). A suitable choice is a 
Pauli-Villars regulator---an additional field with a very large ``mass''
$M'$, whose contribution is added to (\ref{J}) with an opposite sign relative 
to that of $\chi$. If the maximal $k_x$ is some
$\Lambda \gg M'$, the regulator contribution to (\ref{q}) can be computed
analytically. Both the pair production and tunneling are negligible for large $M'$, 
so we replace the regulator mode functions with their WKB expressions, and the 
sum over $k_x$ with an integral, to obtain
\be
q'(t, k_y) = \frac{c_1}{\pi L_y} \tE \; .
\label{q'}
\ee
The resulting expression for $q$ at finite (large) value of $\Lambda$ is then
\be
q_{\Lambda}(t, k_y) = 
2 c_1^2  \sum_{k_x=-\Lambda}^{\Lambda} 
[k_x - \tE(t) ] |{\f}|^2 + q' \; .
\label{regL}
\ee
and the full regularized expression is
\be
q(t, k_y) = \lim_{\Lambda\to\infty} q_{\Lambda}(t, k_y) \; .
\label{reg}
\ee
\section{Numerical Results} \label{sect:num}
As discussed in Sect. \ref{sect:cont}, it is possible to consider geometries
in which the suppression of the order parameter and the driving force on the
vortex are entirely independent functions of time. For example, in the double-arm
geometry of Fig. \ref{fig:dev}, the suppression is determined by the externally
controlled current $I$, while the driving force, by the
biasing flux. Accordingly, we consider here a situation when the order
parameter is suppressed for a relatively long time down to some value that 
allows tunneling, while the driving force exists only for a shorter time: its
biases tunneling and leads to a nonzero value of the average (\ref{J}).

A convenient parametrization of the force $F$ is obtained by defining an effective
current $j'$ related to $E'$ by an effective Maxwell equation
\be
j' = -{1\over 4\pi} \dpar{E'}{t} = -\frac{\dot{E}}{4\pi} +\frac{e F}{\pi\hbar d} \; .
\label{j'}
\ee
For a sufficiently adiabatic pulse, the first term here is much smaller than 
the second, and we neglect it in what follows. According to eq. (\ref{GL}), it is
$j / n_s$, i.e., a quantity akin to the vector potential, that determines how close 
the film is to criticality.
So, we define an effective vector potential $A'$ via the London formula
\be
j' = - \frac{c}{4\pi} \frac{A'}{\bl^2} \; .
\label{A'}
\ee
Combining (\ref{j'}) (with $\dot{E}\approx 0$) and (\ref{A'}), we obtain
\be
F = - \frac{\hbar c d}{8 \alpha_{\rm EM} \bl^2 \xi} \frac{A'}{A_c} \; ,
\label{param}
\ee
where $\alpha_{\rm EM} = 1/137$ is the fine-structure constant, 
and $A_c = \Phi_0/2\pi \xi$ is the critical vector potential.

We consider only biasing pulses that have very small $A'/A_c$ ratios. 
Such pulses do not significantly modify
$\bl$. So, the only difference between $\bl$ and the unperturbed
value of the penetration depth $\lambda$ is due to the broader pulse of the current $I$.
Similarly, the vortex ``mass'' $M(t)$ during the biasing pulse may be assumed constant
and equal to some $M_0$.

For the parameters of the film, we take $d=4$ nm, $\xi = 20$ nm, 
and $\lambda = 150$ nm.
The latter two values take into account the suppression
of $\xi$ and the increase in $\lambda$ due to the small value of the thickness $d$
\cite{Tinkham}. The sizes of the film are $L_x = 1 \mu$m and
$L_y=10 \mu$m.

We assume that the suppression of superconductivity by a pulse of $I$ has reduced $n_s$ 
by a factor of 25. Then, $\bar{\lambda}^2 = 25 \lambda^2$. In our
numerical integrations, we keep $\bl$ and the form of the pulse fixed and scan over 
different values of $M_0$.

The limiting vortex speed can be obtained by estimating the inertial mass of the
vortex $m_v$ and taking the ratio $c_1^2 = \hbar M_0 / m_v$. 
In many cases, the main contribution to $m_v$ comes from the small variation of 
electron density at the vortex core.
In this case, $m_v\sim m k_F d$, and $c_1 \sim v_F$ \cite{Suhl,KL,creep1}. 
However, when we search for
an optimal tunneling path in the Euclidean time, the density at the core becomes
a variational parameter, and it is 
advantageous for it to differ from the average density as little as possible.
In this case, the main contribution to $m_v$ comes from the electric field
produced by the moving vortex, resulting in a much larger 
$c_1 \sim (\xi/\lambda) c$ \cite{Suhl}. For the above values of the parameters, 
we use $c/c_1 = 7.5$.

In what follows, we adopt the system of units in which $c_1 = 1$ and all lengths
are measured in microns. Thus, the unit of time is $1\mu{\rm m}/c_1 = 0.025$ ps.

The easiest way to implement the quantization condition (\ref{quant}) is to
consider $A'$ whose integral over time is zero. We set
\be
\frac{A'}{A_c} = C \frac{t}{t_0} e^{-(t/t_0)^2} 
\ee
and take $C = 0.005$ and $t_0 = 80$. The latter corresponds to 2 ps.

Equation (\ref{eqf}), with
\be
\tE(t) = \frac{cd}{8 \alpha_{\rm EM} \bl^2 \xi} \int_{t_i}^t 
\frac{A'(t')}{A_c} dt'
\ee
and different values of $M(t)= M_0$ was integrated numerically using a 
Runge-Kutta sixth-order integrator. We use $N_x-1$ values of $k_x$:
$k_x L_x / 2\pi = 0, \pm 1, \ldots , \pm (N_x/2 -1)$, with $N_x=32$.

Taking the limit in eq. (\ref{reg}) requires correcting the numerical data
at least by terms of order $M_0^2 / \Lambda^2$. In our case, 
$\Lambda = 30\pi$. The correction was computed by assuming that it dominates the
transport already for $M_0=10$, which is a large enough value to significantly 
suppress tunneling. 

\begin{figure}
\leavevmode\epsfxsize=3in
\epsfbox{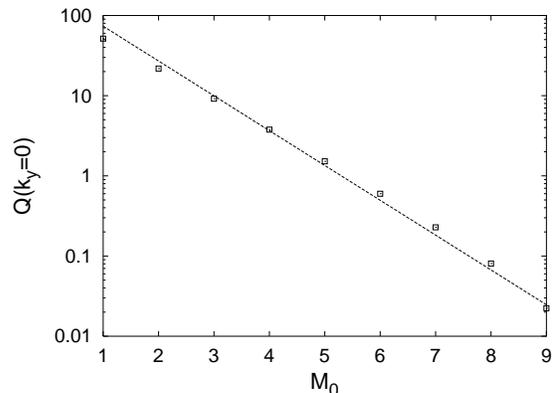}
\caption{Points: total vortex transport (\ref{Q}) for $k_y=0$ and different values
of $M_0$. Line: a const.$\times \exp(-M_0)$ fit.}
\label{fig:cur}
\end{figure}

In Fig. \ref{fig:cur} we plot the total vortex transport
\be
Q(k_y) = L_y \int_{t_i}^{t_f} q_\Lambda(k_y, t) + 0.0215~M_0^2 
\label{Q}
\ee
for $k_y=0$ and several values of $M_0$; $t_{f,i} = \pm 400$. The data are well fit by 
a curve proportional to $\exp(-M_0 L_x)$, which is the instanton exponential.
Note that in this calculation the maximal value
of $\tE$ is $\tE_{\max} \approx 9$, which exceeds the mode spacing 
$2\pi/L_x = 2\pi$. For $\tE \ll 2\pi/L_x$, agreement between the instanton and 
real-time calculations was confirmed analytically in ref. \cite{tran}. We now
confirm the agreement beyond the small $\tE$ limit.

\begin{figure}
\leavevmode\epsfxsize=3in
\epsfbox{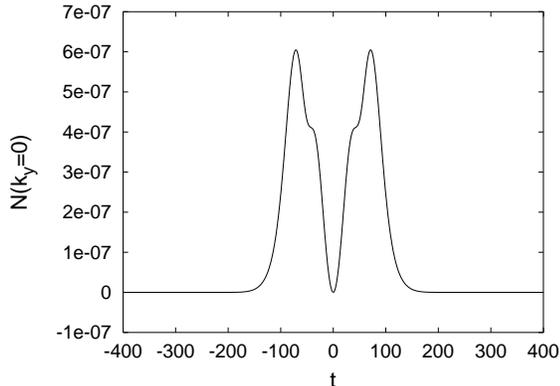}
\caption{Total occupation number for $M_0=5$ and $k_y=0$, as a function of time.}
\label{fig:num}
\end{figure}

Because $t_0 \gg 2\pi / M_0$ for all values of $M_0$ in Fig. \ref{fig:cur},
the transport is to a good accuracy adiabatic. The measure of adiabaticity
is the vortex occupation numbers
\be
n_{\k}(t) = \frac{V}{\omega_{\k}(t)} {\cal E}_{\k}(t) - 1 \; ,
\label{nk}
\ee
where 
\be
{\cal E}_{\k}(t) =  |\dot{\f}|^2 + \omega^2_{\k}(t) |{\f}|^2 \; ,
\label{Ek}
\ee
are the energies (divided by $\hbar$) of the individual modes.
In Fig. \ref{fig:num} we plot the total occupation number
\be
N(k_y, t) = \sum_{k_x=-\Lambda}^{\Lambda} n_{\k}(t) \; ,
\label{N}
\ee
for $k_y = 0$ and $M_0=5$. We see that there is practically no
residual excitation (real vortex-antivortex pairs) in the final state:
at $t=400$, we obtain $N(k_y=0) \sim 10^{-12}$.

\section{Conclusion} \label{sect:concl}
The results of this work are two-fold. First, we have shown that in certain 
geometries (such as the double-arm geometry of Sect. \ref{sect:cont}), it is
possible to independently control the driving force acting on a vortex as
it tunnels across a superconducting film and the suppression of superconductivity 
in the film (i.e., the mass of the vortex). Motivated by this observation, we have
then considered a model problem of vortex tunneling in a toroidal film. The driving
force is modeled by an effective electric field, which biases tunneling and 
results in a nonzero average vortex current.

Second, we have used this model setup to study the dependence of the tunneling rate
on the vortex ``mass'' (more precisely, the pair-production frequency) in
the real-time approach, in which vortices are represented by a fundamental 
quantum field. We have confirmed the
exponential dependence on the ``mass'' found in the instanton (Euclidean) approach.
We have also confirmed that a sufficiently slow, adiabatic variation of the
biasing field can lead to a sizeable vortex current without any real vortex-antivortex
pairs remaining in the final state. This means that induced vortex transport
may be a suitable technique for applications requiring a high-degree of quantum 
coherence, such as quantum memory.


\begin{thebibliography}{10}
\bibitem{1} J. R. Friedman {\emph{et al}}., Nature, {\bf{406}}, 43 (2000).
\bibitem{2} C. H. van der Wal {\emph{et al}}., Science, {\bf{290}}, 773 (2000).
\bibitem{tipping} X. Zhou {\em et al.}, IEEE Trans. 
Appl. Supercond. {\bf 11}, 1018 (2001) [quant-ph/0102090].
\bibitem{deex} M. Crogan, S. Khlebnikov, and G. Sadiek,
Supercond. Sci. Technol. {\bf 15}, 8 (2002) [quant-ph/0105038]. 
\bibitem{Fuchs} K. Fuchs, Proc. Cambridge Phil. Soc. {\bf 34}, 100 (1938).
\bibitem{KK} N. B. Kopnin and V. E. Kravtsov, 
Pis'ma ZhETF {\bf 23}, 631 (1976) [JETP Lett. {\bf 23}, 578 (1976)].
\bibitem{Volovik} G. E. Volovik, Zh. Eksp. Teor. Fiz.
{\bf 104}, 3070 (1993) [JETP {\bf 77}, 435 (1993)].
\bibitem{micro} A. van Otterlo, M. Feigel'man, V. Geshkenbein, and G. Blatter,
Phys. Rev. Lett. {\bf 75}, 3736 (1995).
\bibitem{FGLV} M. V. Feigel'man, V. B. Geshkenbein, A. I. Larkin, and 
V. M. Vinokur, Pis'ma ZhETF {\bf 62}, 811 (1995) [JETP Lett. {\bf 62}, 834 (1995)].
\bibitem{tran} S. Khlebnikov, quant-ph/0210019.
\bibitem{Pearl} J. Pearl, Appl. Phys. Lett. {\bf 5}, 65 (1964).
\bibitem{GF} L. I. Glazman and N. Ya. Fogel, Fiz. Nizk. Temp. {\bf 10}, 
95 (1984) [Sov. J. Low Temp. Phys. {\bf 10}, 51 (1984)].
\bibitem{Ao} P. Ao, J. Low Temp. Phys. {\bf 89}, 543 (1992).
\bibitem{Tinkham} M. Tinkham, Phys. Rev. {\bf 110}, 26 (1958).
\bibitem{Suhl} H. Suhl, Phys. Rev. Lett. {\bf 14}, 226 (1965).
\bibitem{KL} M. Yu. Kupriyanov and K. K. Likharev, ZhETF {\bf 68}. 1506 (1975)
[Sov. Phys. JETP {\bf 41}, 755 (1975)].
\bibitem{creep1} G. Blatter, V. B. Geshkenbein, and V. M. Vinokur,
Phys. Rev. Lett. {\bf 66}, 3297 (1991).
\end{thebibliography}
\end{document}